\documentstyle[12pt,aasms4]{article} 
\begin{document}

\title{Cosmic Microwave Background Radiation in the Direction of a Moving
Cluster of Galaxies with Hot Gas: Relativistic Corrections} 

\author{S.Y. Sazonov\altaffilmark{1,2} and R.A. Sunyaev\altaffilmark{1,2}}
\altaffiltext{1}{MPI f\"ur Astrophysik,
Karl-Schwarzschild-Str.1, 86740 Garching bei M\"unchen, Germany}
\altaffiltext{2}{Space Research Institute (IKI), Profsouznaya 84/32, Moscow
117810, Russia}

\begin{abstract}
It has been recently realized (Rephaeli 1995) that the relativistic
corrections to the spectral distortions of the cosmic microwave background
(CMB) measured in the direction of clusters of galaxies containing hot gas
are significant and should be detectable with the forthcoming experiments.
In the present paper we calculate the correction terms that are proportional
to $V_{\rm r}/c\times kT_e/m_ec^2$ and $(V/c)^2$ to the standard formulae
describing the spectral distortions caused by the bulk motion of the free
electrons (kinematic effect) and due to the presence of the hot gas (thermal
effect) for the case of a cluster having a peculiar velocity $V$ ($V_{\rm
r}$ is its radial component). The results of our analytical calculations are
confirmed by Monte-Carlo simulations (Sazonov \& Sunyaev 1998). 
\end{abstract}


\section{Introduction}
Thomson scattering of the cosmic microwave background (CMB) radiation by hot
electrons in the intergalactic gas in clusters of galaxies modifies the
spectrum of the CMB (Sunyaev and Zel'dovich 1972). Zel'dovich and Sunyaev
(1969), basing on the Kompaneets equation (1957), derived a simple
formula describing the spectral form of the distortion, which is
proportional to the parameter $y=(kT_e/m_ec^2) \tau$, where $\tau$ is the
Thomson optical depth along the line of sight. The effect has now been
observed from a number of clusters of galaxies (see Birkinshaw 1998 for
review).

Recently, interest to this effect has been
reactivated in view of the perspectives of accurate measurement of
the CMB distortions in a number of experiments, both ground-based and on
balloons, by the MAP spacecraft and especially by the Planck Surveyor mission
scheduled to be flown in the middle of the next decade. These 
activities were motivated by the fact that the gas temperature is so high in
the clusters of galaxies (ranging between 3 and 17~keV, Tucker
et al. 1998) that the scattering electrons have thermal velocities of the
order of 0.1 -- 0.3 $c$, so one has to include into consideration the 
relativistic corrections to obtain an accurate result. Rephaeli (1995),
basing on extensive previous work (Wright 1979, Fabbri 1981, Taylor \&
Wright 1989, Loeb et al. 1991) has demonstrated by means of numerical
calculations the relevance of the relativistic corrections for the future
experiments. Stebbins (1997), Itoh et al. (1998), and Challinor
\& Lasenby (1998) used a Fokker-Planck approximation of the relativistic
photon kinetic equation to obtain corrections, written as series in powers
of $kT_e/m_ec^2$, to the standard nonrelativistic solution. These results
have proved to be in excellent agreement with those of Rephaeli,
demonstrating the applicability of the diffusion approximation to the
problem at hand, despite the small optical depths of the clusters of
galaxies ($\tau \sim 0.01$).

A gas cloud moving rapidly relative to the CMB along the observer's line of
sight must significantly modify the spectrum of the CMB in addition to the  
thermal effect. The change in the brightness temperature caused by this
``kinematic'' effect is to first order simply proportional to the radial
component of the cluster velocity $\sim (V_{\rm r}/c)\tau$ (Sunyaev \&
Zel'dovich 1980). The effect should be detectable in the future, and will
enable measurement of cluster peculiar velocities, with significant
implications for studies of the large-scale structure of 
the universe. It is obvious that corrections similar to those found for
the thermal effect must exist and should be taken into account for the
kinematic effect, if one wants to find the correct solution for the case
of a moving cluster. In this paper we calculate the next-order changes in
the spectrum of the CMB related to the cluster peculiar 
velocity by solving the photon kinetic equation. We have obtained simple
formulae giving the correction terms of the orders of $V_{\rm r}/c\times
kT_e/m_ec^2$ and $(V_{\rm r}/c)^2$. Our method is similar to that used by
Psaltis \& Lamb (1997) who considered the more general problem of
comptonization in a moving media. The solution these authors have obtained,
although applicable to many astrophysical situations, does not contain the
$O(V_{\rm r}/c\times kT_e/m_ec^2)$ term, because this term is third-order in
electron velocity, whereas their solution is accurate only to second order
in it. We aslo confirm the existence of the term of order $(kT_e/m_ec^2)^2$
found earlier using techniques different from ours (Rephaeli 1995, Stebbins
1997, Itoh et al. 1998, and Challinor \& Lasenby 1998). Earlier we have
found all the correction terms mentioned above numerically using Monte-Carlo
simulations (Sazonov \& Sunyaev 1998).

\section{Scattering of the CMB by a directed beam of electrons} 

From the point of view of the observer each electron in the intergalactic
gas scatters the CMB photons independently. We can therefore first consider the
problem of scattering of the CMB by a directed beam of monoenergetic
electrons having a density $N_{\rm e}$ and moving at velocity $\bf v$.
Once we have found a solution to this simplified problem, we will be able to
consider the more general problem of scattering of the CMB by a cloud of
thermal electrons having a peculiar motion, which corresponds to the real
situation of a cluster of galaxies, by simply averaging the result obtained
for the directed beam over a drifting Maxwellian distribution of electron
velocities. We will use two coordinate frames. Quantities with subscript 0
refer to the system where the electrons are at rest, while quantities
without subscript refer to the frame that is fixed to the CMB, hereafter
referred to as the laboratory frame. It is easy to show that our results are
valid for a cluster at any redshift (see Sazonov \& Sunyaev 1998). In the
laboratory frame the initial occupation number in the photon phase space is
planckian with a temperature $T_{\rm CMB}$: $n =1/(e^x-1)$, where
$x=h\nu/kT_{\rm CMB}$. The corresponding spectral intensity is

\begin{equation}
I_{\nu}= \frac{2(kT_{\rm CMB})^3}{(hc)^2} \frac {x^3}{e^x-1}
\label{eq:ix}
\end{equation}

The occupation number is invariant with respect to the Lorentz
transformations of the frequency and direction of motion of a photon

\begin{equation}
\nu_0=\frac{\nu}{\gamma(1+\beta\mu_0)}=\gamma (1-\beta\mu) \mbox{ and } 
\mu_0=\frac{\mu-\beta}{1-\beta\mu},
\label{eq:lorentz}
\end{equation}
where $\beta=v/c$, $\gamma=(1-\beta^2)^{-1/2}$, $\mu=\cos{\theta}$, and
$\theta$ is the angle between $\bf v$ and the photon velocity (Landau \&
Lifshitz 1975). Therefore, in the electron rest frame the occupation number
depends on the photon incident direction
 
\begin{equation}
n_0 =\frac{1}{e^{x_0 \gamma (1+\beta\mu_0)}-1},
\end{equation}

Ignoring induced scattering and the change in the photon frequency in each
scattering due to the recoil effect, we can write down the photon kinetic
equation in the electron rest frame as follows: 

\begin{equation}
\frac{d}{dt_0} n_0(\mu_0,\nu_0)= c N_{{\rm e}0} \int \frac
{d\sigma}{d\Omega^\prime_0}
[n_0(\mu^\prime_0,\nu_0)-n_0(\mu_0,\nu_0)]\,d\Omega^\prime_0, 
\end{equation}
Integrating the Thomson differential cross-section over the azimuthal angle
(Chandrasekhar 1950) one derives

\begin{equation}
\frac{d}{dt_0} n_0(\mu_0,\nu_0)= \frac{3 c N_{{\rm e}0} \sigma_{\rm T}}
{16\pi} \int_{-1}^{1} (3+3\mu^2_0\mu^{\prime 2}_0-\mu^2_0-\mu^{\prime
2}_0) [n_0(\mu^\prime_0,\nu_0)-n_0(\mu_0,\nu_0)]\,d\mu^\prime_0 
\label{eq:bolz0}
\end{equation}

We now evaluate the collision integral in equation (\ref{eq:bolz0}) by   
expanding $n_0$ up to forth order in $\beta$. The integral is
then easily taken, yielding:

\begin{eqnarray}
\frac {1}{n_0(\mu_0,\nu_0)} \frac{dn_0(\mu_0,\nu_0)}{dt_0}= 
c N_{{\rm e}0}\sigma_{\rm T} \left\{\frac{x_0 e^{x_0}}{e^{x_0}-1}\, 
\left[\mu_0 \beta+\frac{3(-1+3\mu^2_0)}{20} x_0\beta^2+
\frac{\mu_0}{2}\beta^3+\frac{\mu_0}{2}x_0\beta^3 \right.\right.
\nonumber \\
\left.+\frac{\mu^3_0}{6}x_0^2\beta^3+
\frac{3(-1+3\mu^2_0)}{20}x_0\beta^4+
\frac{3(-1+3\mu^2_0)}{40}x_0^2\beta^4+
\frac{-6-3\mu^2_0+35\mu^4_0}{840}x_0^3\beta^4\right]
\nonumber\\
+\left(\frac{x_0 e^{x_0}}{e^{x_0}-1}\right)^2\,
\left[\frac{3+\mu^2_0}{10}\beta^2-
\frac{\mu_0}{2}\beta^3-\frac{\mu_0(3+\mu^2_0)}{20}x_0\beta^3+
\frac{3+\mu^2_0}{10}\beta^4+\frac{3-\mu^2_0}{8}x_0\beta^4 \right.
\nonumber \\
\left.+\frac{4-\mu^2_0-\mu^4_0}{40}x_0^2\beta^4 \right]+
\left(\frac{x_0 e^{x_0}}{e^{x_0}-1}\right)^3\,
\left[\frac{\mu_0(3+\mu^2_0)}{10}\beta^3-\frac{3+\mu^2_0}{10}\beta^4 \right.
\nonumber \\
\left.\left.+\frac{-36+3\mu^2_0+7\mu^4_0}{140}x_0\beta^4 \right]+
\left(\frac{x_0
e^{x_0}}{e^{x_0}-1}\right)^4\,\left[\frac{3(2+\mu^2_0)}{35}\beta^4\right]
\right\},
\label{eq:rest}
\end{eqnarray}

Our next step is to calculate the corresponding scattering rate as measured
in the laboratory frame. This can be done by making use of the
Lorentz-invariance property of the photon occupation number. It is
easily shown (see e.g. Peebles 1971) that

\begin{equation}
\frac{dn(\mu,\nu)}{dt}= \frac{1}{\gamma(1+\beta\mu_0)}
\frac{dn_0(\mu_0,\nu_0)}{dt_0} 
\label{eq:rate}
\end{equation}

Using relations (\ref{eq:lorentz}), (\ref{eq:rate}) and substituting
$N_{e0}=N_e/\gamma$ (due to the Lorentz-transformation of volume) for the
electron density, we derive from equation (\ref{eq:rest}) 

\begin{eqnarray}
\frac{1}{n(\mu,\nu)} \frac{dn(\mu,\nu)}{dt}= c N_{\rm e}\sigma_{\rm T}
\frac{x e^x}{e^x-1} \left\{\beta\mu+
\beta^2\,\left[-1-\mu^2+\frac{3+11\mu^2}{20}F\right]+
\beta^3\mu\,\left[2-\frac{31+11\mu^2}{20}F \right.\right.
\nonumber\\
\left.+\frac{9+13\mu^2}{120}(2F^2+G^2)\right]+
\beta^4\,\left[-1-\mu^2+\frac{17+53\mu^2}{20}F-
\frac{9+66\mu^2+13\mu^4}{120}(2F^2+G^2) \right.
\nonumber\\
\left.\left.+\frac{3+33\mu^2+28\mu^4}{420}F(F^2+2G^2)\right]\right\},
\label{eq:lab}
\end{eqnarray}
where $F=x\coth{(x/2)}$, and $G=x/\sinh{(x/2)}$.

Compton scattering must save the total number of photons. We have verified that
all the $\beta$ terms in equation (\ref{eq:lab}) indeed vanish after the
integration over photon direction and frequency: $d/dt \int n \nu^2 \,d\nu
\,d\mu=0$.

Another known integral property of the process of compton scattering is the
energy exchange rate between an electron and an isotropical radiation field.
The radiation energy density $\epsilon_{\rm r}$ (see e.g. Pozdnyakov et 
al. 1983) should increase with time as

\begin{equation}
\frac {d\epsilon_{\rm r}}{dt} = \frac{4}{3} c N_e \sigma_{\rm
T}\epsilon_{\rm r} (\gamma^2-1)
\label{eq:transfer}
\end{equation}
Taking an integral $d/ dt \int n \,\nu^3 \,d\nu\,d\mu$ we find in our case

\begin{equation}
\frac {d\epsilon_{\rm r}}{dt} = \frac{4}{3} c N_e \sigma_{\rm
T}\epsilon_{\rm r} (\beta^2+\beta^4),
\label{eq:transfer1}
\end{equation}
which is identical to dependence (\ref{eq:transfer}) to forth order in
$\beta$. 

\section{Scattering of the CMB by the hot gas in a moving cluster}

Consider now a cluster of galaxies moving with a peculiar velocity $V$ at an
angle $\theta$ ($\mu=\cos{\theta}$) relative to the vector drawn from the
cluster to the observer. The cluster contains hot gas, and the distribution of
the electrons in the cluster rest frame is assumed to be relativistic
Maxwellian with a temperature $T_e$: $dN_e=A \exp{[-E_0({\bf p}_0)
m_ec^2/kT_e^2]} \,d{\bf p}_0$, where ${\bf p}_0$ is the electron momentum,
$E_0$ is the electron energy, and $A$ is the normalization constant. The
corresponding distribution in the laboratory frame is obtained via the
Lorentz-transformation of ${\bf p}$ 

\begin{equation}
p_x=\gamma (p_{x0}+\frac{V}{c^2} E_0)\mbox{; } p_y=p_{y0}\mbox{; } p_z=p_{z0},
\end{equation}
where $\gamma=(1-V^2/c^2)^{-1/2}$, and axis $X$ is drawn along the direction
of the cluster peculiar motion (Landau \& Lifshitz 1975). We can average
equation (\ref{eq:lab}) over the resulting electron velocity distribution
expanded in powers of $V/c$ and $T_e$ to get

\begin{eqnarray}
\frac{\delta n(\nu)}{\tau n(\nu)}=
\frac{xe^x}{e^x-1}\,\left\{\frac{V}{c}\mu+\frac{kT_e}{m_ec^2}(-4+F)
+\left(\frac{V}{c}\right)^2\,\left[-1-\mu^2+\frac{3+11\mu^2}{20}F\right]\right.
\nonumber\\
+\frac{V}{c}\frac{kT_e}{m_ec^2}\mu\,\left[10-\frac{47}{5}F
+\frac{7}{10}(2F^2+G^2)\right]
\nonumber\\
\left.+\left(\frac{kT_e}{m_ec^2}\right)^2\left[-10+\frac{47}{2}
F-\frac{42}{5}F^2+ \frac{7}{10}F^3+\frac{7}{5}G^2(-3+F)\right]\right\},
\label{eq:final}
\end{eqnarray}
Here we have replaced an integral over time $\int N_e(r)\sigma_{\rm T}
c\,dt$ by an integral along the line of sight $\tau=\int N_e(r)\sigma_{\rm
T} \,dr$, where $\tau \ll 1$. A Monte-Carlo computation that takes into
account only single-scattering events proves the validity of this transition
(see a detailed discussion in Sazonov \& Sunyaev 1998).   
 
The subsequent terms in equation (\ref{eq:final}) physically correspond to
increasing orders in $\beta$ in equation (\ref{eq:lab}), which is in the
current problem a sum of the electron thermal and peculiar velocities. The
first term (of order $V/c$) describes the kinematic 
effect. The second term, which is proportional
to the second power of the thermal velocity, describes the thermal effect.
The $O[(V/c)^2]$ term is the relativistic 
correction to the kinematic effect for a cloud of cold electrons. The
``interference'' term that is proportional to $V/c\times kT_e/m_ec^2$ draws
from the term of order $\beta^3$ in equation ({\ref{eq:lab}). We have
ascertained its existence by means of Monte-Carlo simulations
(Sazonov \& Sunyaev 1998). This term constitutes the leading relativistic
correction to the kinematic effect and is the main subject of the present
paper. Finally, the term of order $(kT_e/m_ec^2)^2$ (forth-order in $\beta$)
is the relativistic correction to the thermal effect found earlier using a
Fokker-Planck approximation (Stebbins 1997, Itoh et al. 1998, Challinor and
Lasenby 1998).

\section{Properties of the CMB Spectral Distortion}

The distortion of the spectral intensity of the CMB is related to the
corresponding change in the photon occupation number by the equation

\begin{equation}
\delta I_{\nu}=I_{\nu} \frac{\delta n(\nu)}{n(\nu)}
\label{eq:ix},
\end{equation}

In Fig.~1 we plot this distortion at $kT_e/m_ec^2=0.02$,
$V/c=0.01$ (the large value for the peculiar velocity has been chosen for
illustration purposes), for two opposite directions of the cluster motion
$\mu=1$ and $\mu=-1$. One can see that the contribution from the newly
found term $O(V/c\times kT_e/m_ec^2)$ to the total effect is significant. This
contribution reaches its maximum at $x=3.34$, i.e. near the frequency
$x_{\rm c}\sim 3.83$ at which the thermal effect vanishes and where
measurements of the kinematic effect would seem most promising.

In our previous paper (Sazonov \& Sunyaev 1998) we have performed
Monte-Carlo simulations to determine the spectral changes in the CMB  for
various sets of parameters. As evident from Fig.~1, which uses the result of
that paper, the correctness of the analytical formula (\ref{eq:final}) is
confirmed by the numerical calculations. 

Using equation (\ref{eq:final}) we have obtained a simple approximation
formula describing the position of the crossover frequency ($X_0$), i.e. the
frequency at which the distortion of the incident microwave spectrum is zero

\begin{equation}
X_0=3.830 \,\left(1-0.31 \frac{V_{\rm r}}{c} \frac{m_ec^2}{kT_e}
+1.1 \frac{kT_e}{m_ec^2}-0.6 \frac{V_{\rm r}}{c}\right)
\label{eq:cross}
\end{equation}
The correction term of order $kT_e/m_ec^2$ in this formula was known before
our work (Rephaeli 1995, Itoh et al. 1998, Challinor \& Lasenby 1998). The
$O(V_{\rm}/c\times m_ec^2/kT_e)$ term is due to the kinematic effect.

We have obtained similar approximation formulae for the positions ($X_{min}$
and $X_{max}$) and values ($J_{min}$ and $J_{max}$) of the minimum and
maximum of the spectral dependences shown in Fig.~1. They are as follows

\begin{equation}
X_{min}=2.266 \,\left(1-0.23 \frac{V_{\rm r}}{c} \frac{m_ec^2}{kT_e}
-0.1 \frac{kT_e}{m_ec^2} +1.4 \frac{V_{\rm r}}{c}\right),
\end{equation}

\begin{equation}
J_{min}=-2.059\, \frac{2(kT_{\rm CMB})^3}{(hc)^2} \frac{kT_e}{m_ec^2}
\tau\,\left(1-0.76\frac{V_{\rm r}}{c} \frac{m_ec^2}{kT_e}-3.4
\frac{kT_e}{m_ec^2} +0.3 \frac{V_{\rm r}}{c}\right),  
\end{equation}

\begin{equation}
X_{max}=6.511 \,\left(1-0.09 \frac{V_{\rm r}}{c} \frac{m_ec^2}{kT_e}+2.5
\frac{kT_e}{m_ec^2} -0.5 \frac{V_{\rm r}}{c}\right), 
\end{equation}

\begin{equation}
J_{max}=3.390 \, \frac{2(kT_{\rm CMB})^3}{(hc)^2} \frac{kT_e}{m_ec^2} \tau
\,\left(1+0.43\frac{V_{\rm r}}{c} \frac{m_ec^2}{kT_e}-6.2
\frac{kT_e}{m_ec^2} +0.6 \frac{V_{\rm r}}{c}\right), 
\end{equation}

Finally, one can calculate the excess in the CMB energy flux in the
direction of the cluster ($\delta I= \int \delta I_{\nu}\,d\nu$)   

\begin{equation}
\delta I= \tau I \,\left[4\mu\frac{V}{c}+4\frac{kT_e}{m_ec^2}+
+(7\mu^2-1)\left(\frac{V}{c}\right)^2+20\mu\frac{V}{c}\frac{kT_e}{m_ec^2}
+10\left(\frac{kT_e}{m_ec^2}\right)^2\right], 
\label{eq:lum}
\end{equation}
where $I=b T^4_{\rm CMB}/(4\pi c)$, and $b$ is the Stefan-Boltzmann constant.

One can see again that the relativistic correction of order $V/c\times
kT_e/m_ec^2$ is important. Equation (\ref{eq:lum}) allows one to calculate
the rate of energy exchange between the hot gas and the CMB. We immediately
see that the $O(kT_e/m_ec^2)$ and $O[(kT_e/m_ec^2)^2]$ terms are the
two leading terms in the relativistic formula that gives the energy transfer
rate averaged over a Maxwellian distribution of electron velocities:
$d\epsilon_{\rm r}/dt=4/3\, c N_e \sigma_{\rm T} \epsilon_{\rm r} \langle
\gamma^2-1 \rangle$. Integration of the term of order $(V/c)^2$ in formula
(\ref{eq:lum}) over the observing angle ($\mu$) leads again to relation
(\ref{eq:transfer1}) for the energy transfer rate due to the bulk motion of the
electrons, as one should have expected.

We have also obtained approximation formulae giving separately the flux
from the ``negative'' source, i.e. integrated over the frequency range
$(0,X_0)$ and that from the positive source (integrated from $X_0$ to
$\infty$), taking into account dependence (\ref{eq:cross})

\begin{equation}
\delta I_{-} = \tau I \left[
-1.35\frac{kT_e}{m_ec^2}+1.5\frac{V_{\rm r}}{c}-0.4 \left(\frac{V_{\rm
r}}{c}\right)^2 \frac{m_ec^2}{kT_e}
+3.3\left(\frac{kT_e}{m_ec^2}\right)^2-2.4 \frac{V_{\rm r}}{c}
\frac{kT_e}{m_ec^2}\right]
\end{equation}

\begin{equation}
\delta I_{+} = \tau I \left[
5.35\frac{kT_e}{m_ec^2}+2.5\frac{V_{\rm r}}{c}+0.4 \left(\frac{V_{\rm
r}}{c}\right)^2 \frac{m_ec^2}{kT_e}
+6.7\left(\frac{kT_e}{m_ec^2}\right)^2+22.4 \frac{V_{\rm r}}{c}
\frac{kT_e}{m_ec^2}\right]
\end{equation}

\acknowledgements
The authors wish to thank Dr. Eugene Churazov for valuable discussions. 
We are grateful to Prof. Itoh for sending us the information about a
paper (astro-ph/9804051 at http://xxx.lanl.gov/) where results similar to ours
are obtained using a Fokker-Planck approximation.


\clearpage

\figcaption{(a) An example of the CMB spectral distortion (in units of
$2(kT_e)^3/(hc)^2)$ due to the Sunyaev-Zel'dovich effect for the following
set of cluster parameters: $kT_e/m_ec^2=0.02$ ($T_e=10.2$~keV), $V/c=0.01$
($V=3000$~km/s), and $\mu=1$ (the cluster moves toward us). The
solid line shows the cumulative effect, which was calculated by summing the
analytical expressions for terms of different orders in $kT_e/m_ec^2$ and
$V/c$ as given in Itoh et al. (1998) and the present paper. The
contributions from the following components are shown: $O(kT_e/m_ec^2)$
(dotted curve), $O[(kT_e/m_ec^2)]^2$ (short-dashed curve), $O(V/c)$
(long-dashed curve), $O[(V/c)^2]$ (dash-dotted curve), and $O(V/c\times
kT_e/m_ec^2)$ (long-short dashed curve). For comparison, the result of
Monte-Carlo simulations is shown as a histogram. (b) Same as (a), but the
cluster moves outwards from the observer ($\mu=-1$). For comparison, the
result shown in Fig.~1 is repeated. 
}  
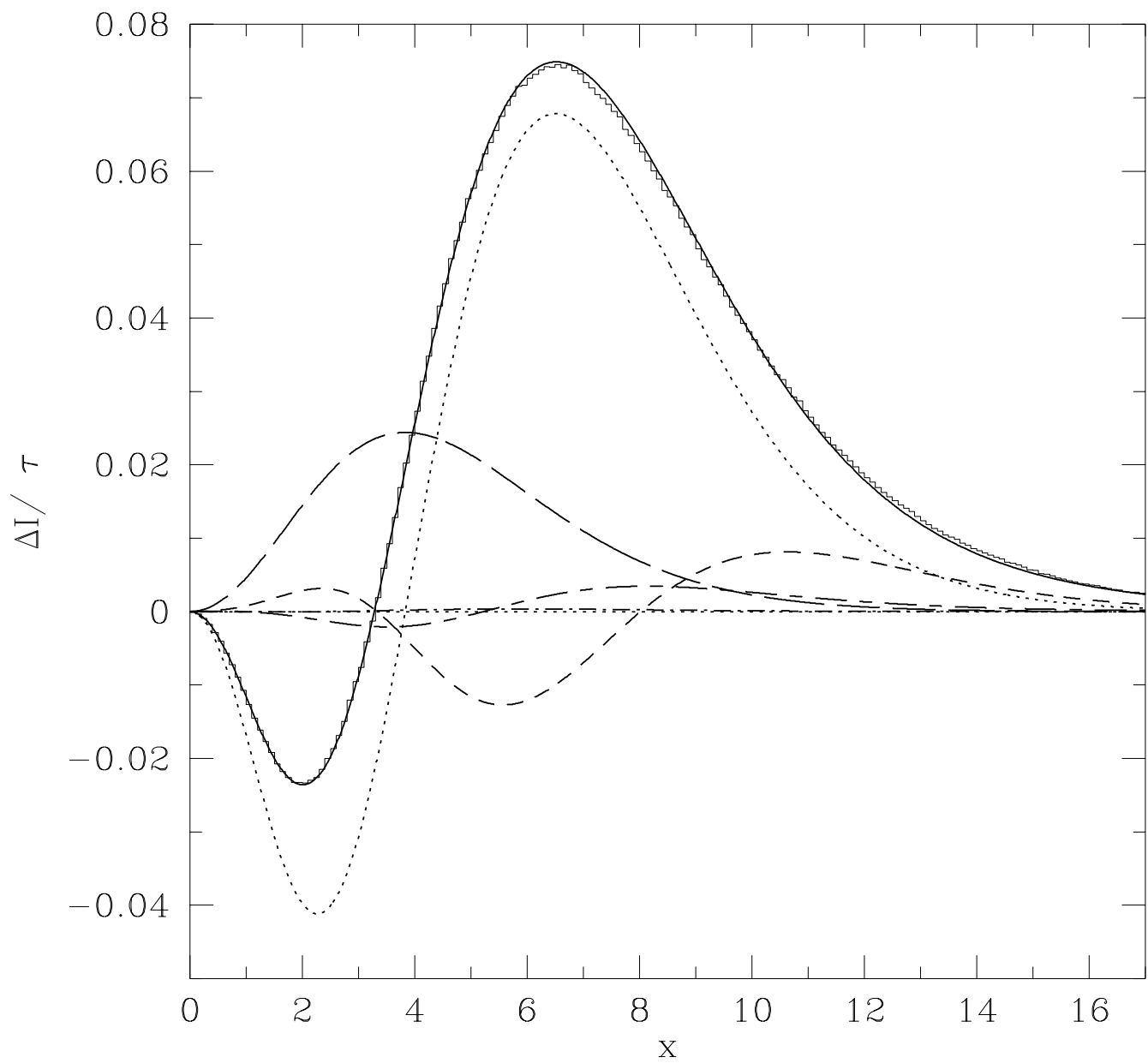
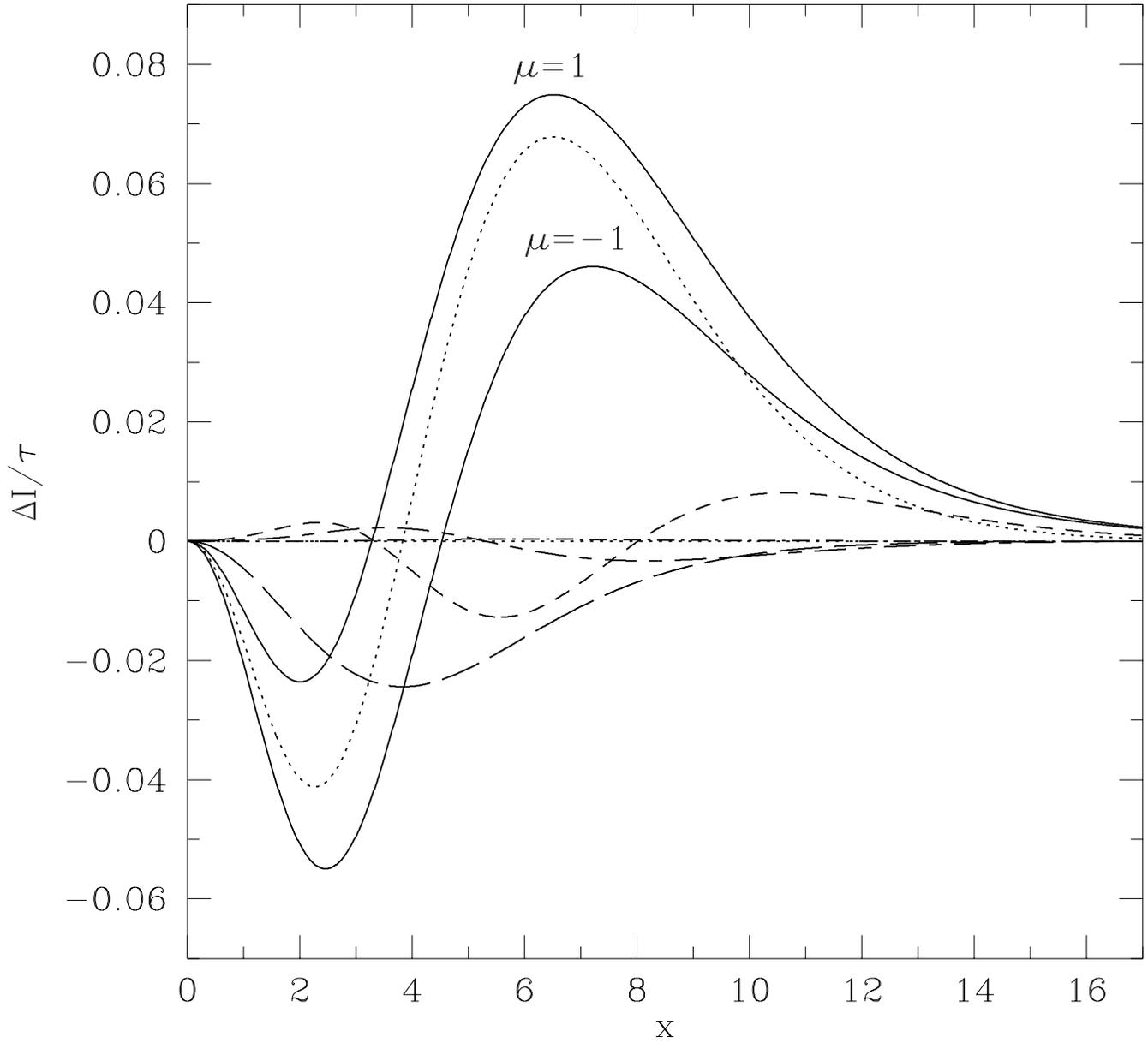
\end{document}